\documentclass[runningheads]{llncs}

\usepackage{graphicx}

\graphicspath{{img/}}

\usepackage{multirow}

\AtBeginDocument{%
  \providecommand\BibTeX{{%
    \normalfont B\kern-0.5em{\scshape i\kern-0.25em b}\kern-0.8em\TeX}}}

\begin{document}

\title{Exploring Workspace Awareness Needs during Mixed-Presence Collaboration on Interactive Wall-Sized Displays: Results from a Focus Group}

\titlerunning{Workspace Awareness Needs during Mixed-Presence Collaboration on WSDs}



\author{Adrien Coppens\inst{1}\orcidID{0000-0002-2841-6708} 
	\and Lou Schwartz\inst{1}\orcidID{0000-0002-8645-5326}
	\and Valerie Maquil\inst{1}\orcidID{0000-0002-0198-3729}
}
%
\authorrunning{A. Coppens et al.}
%
\institute{Luxembourg Institute of Science and Technology, Esch-sur-Alzette, Luxembourg
	\email{firstname.lastname@list.lu}}

\maketitle              


\begin{abstract}
To enhance workspace awareness for mixed-presence meetings with large displays, previous work propose digital cues to share gestures, gaze, or entire postures. While such cues were demonstrated useful in horizontal or smaller workspaces, efforts have focused on isolated elements in controlled settings. It is unknown what needs would emerge with a more realistic setting and how they could be addressed with workspace awareness cues.
In this paper, we report on the results of a focus group, centred around users' perceptions while testing a mixed-presence scenario on wall-sized displays. We analyse the gathered comments using Gutwin and Greenberg’s workspace awareness framework to identify the most relevant needs. Our results lead to a refinement of the original framework for wall-sized displays and in particular to a categorization into three types of workspace awareness components (i) the Environment, (ii) Actions and (iii) Attention.
\end{abstract}



\begin{figure}
    \centering
  \includegraphics[width=\textwidth]{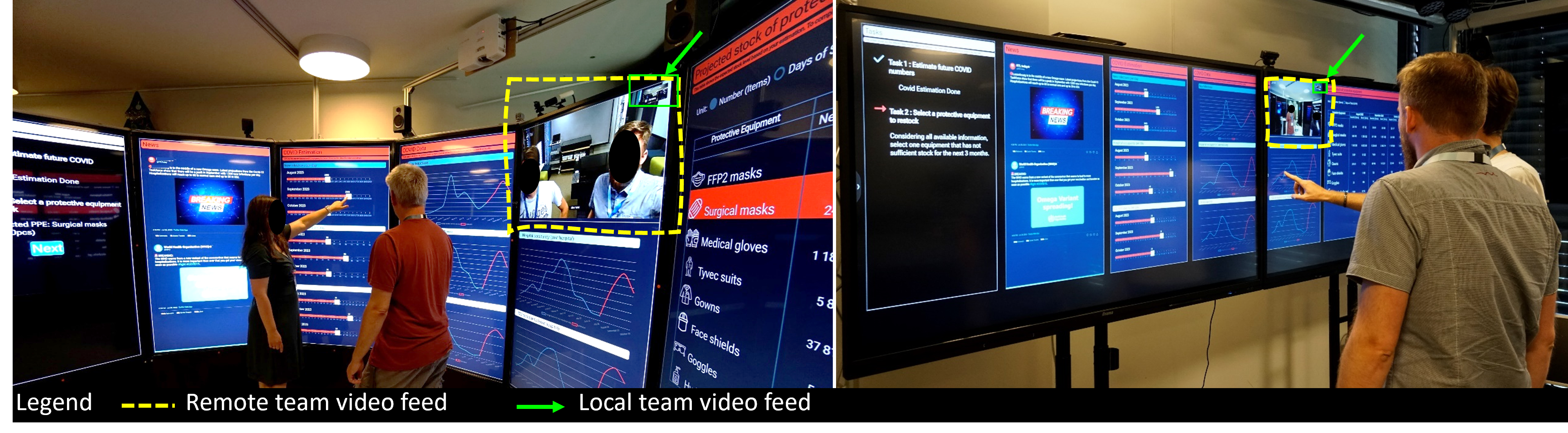}
  \caption{Both locations of the mixed-presence collaboration setup used for the user study. 
  On the left, the 
  Immersive Arena. On the right, the flat wall-sized display.
  The yellow dotted line shows the remote video feed; 
  the green solid line and 
  the arrow show the local video feed.}
  \label{fig:teaser}
\end{figure}


\section{Introduction}
Interactive wall-sized displays (WSDs) provide large benefits for information visualisation and visual data analysis~\cite{langner2019multiple}. Their extensive size combined with their resolution enables the presentation of large amounts of data and supports users in gaining more insights about the data~\cite{rajabiyazdi2015understanding}.
WSDs further facilitate collocated collaboration as group members can observe each other and rely on non-verbal communication such as postures, gestures, gaze direction, and facial expressions.

However, when seeking to limit travel costs or reduce the carbon footprint, alternatives to face-to-face collaboration need to be found.

Videoconferencing tools are often used in such contexts but rely mostly on audio-video links that strongly limit the acquisition of non-verbal awareness information. 
This issue is particularly relevant in the context of decision-making with WSDs, as collaborators are naturally making use of many body movements and hand gestures~\cite{langner2019multiple,anastasiou2020you}.
With conventional audio-video based support, these are not accessible remotely, which leads to difficulties in communication and engagement~\cite{fullwood2006effect}.


Previous work suggest to mitigate that issue through visual cues such as pointers, sketches, or hand gestures. However, most of these cues were designed for smaller or different workspaces (e.g. tabletops~\cite{yamashita2011improving}), targeting dyads (with one person on each site)~\cite{edelmann2013preserving}, or using head-mounted mixed-reality technology~\cite{bai2020user}.

In this paper, we explore the awareness problems that users encounter during a realistic scenario involving distant collaborators on WSDs with basic remote collaboration support. Our aim is to identify which previous observations apply in this context and which additional needs would emerge. 
We first report on related work, then describe the study consisting in a focus group session around the role-playof a mixed-presence (i.e. partially remote) decision-making scenario on WSDs and a brainstorming around encountered problems. We then present our results based on the gathered comments structured using 
Gutwin et al.'s workspace awareness (WA) framework~\cite{gutwin2002descriptive} to propose an adjusted WA framework dedicated to mixed-presence collaboration using WSDs.

\section{Related work}\label{sec:SOTA}
The splitting of teams accross two locations is often referred to as ``mixed-presence'' collaboration, with \textit{``both collocated and distributed participants working over a shared visual workspace in real-time''}~\cite{tang2003display}.

In mixed-presence settings, effective collaboration requires the sharing of essential information regarding the activities of other group members.
Gutwin et al. define the related concept of workspace awareness (WA) as ``\textit{the up-to-the-moment understanding of another person’s interaction with a shared workspace}''~\cite{gutwin2002descriptive}. 
WA deals with the presence and location of collaborators, with the actions they are undertaking and the objects they are acting on, but also with where their attention is focused.
To transmit non-verbal awareness information to remote collaborators, mixed-presence groupware designers typically rely on, often visual, cues.
Since such cues convey WA information, we call them WA cues.

They provide users with important awareness information that make their collaborators’ actions and intentions clear, which leads to seamlessly aligned and integrated activities and pushes them to assist each other ~\cite{yuill2012mechanisms}. This in turn helps the collaboration process, by facilitating coordination and making it both less effortful and less error-prone~\cite{niemantsverdriet2019designing}.


Overall, the existing literature on mixed-presence collaboration has mostly focused on traditional desktop systems (e.g. sharing mouse cursor positions~\cite{osawa2006aggregate}) and more recently on augmented and virtual reality-based collaboration~\cite{bai2020user,piumsomboon2019effects}, including questions related to the combination of WA cues~\cite{kim2019evaluating} in that context. 
Large interactive displays have received limited attention in that regard, with most previous work focusing on tabletops~\cite{yamashita2011improving} or targeting dyads (with one person on each site)~\cite{edelmann2013preserving}.
This leads to solutions that may not apply to WSDs or scale well with more users, e.g. full body representations of remote collaborators~\cite{kuechler2010collaboard}.

Moreover, the scarce literature regarding the particular context of WSDs has mostly focused on isolated awareness elements and cues, and a more holistic view on the topic is required.
To identify the best techniques to overcome the presented challenges, more research is needed to better understand requirements and opportunities regarding WA in mixed-presence collaboration across WSDs.

\section{Methodology and process} \label{sec:study}
In this paper, we focus on mixed-presence collaborative decision-making on WSDs. 
We are interested in the following research questions: \textbf{(RQ1)} which awareness problems and needs do users perceive during mixed-presence meetings with WSDs, and \textbf{(RQ2)} how to describe workspace awareness information needed during mixed-presence collaboration on WSDs?

To answer our research questions, we organised a focus group~\cite{cyr2019focus}, aiming to identify collaboration problems and generate ideas for better WA support. 
After welcoming participants and introducing them to the research and its objectives, we gathered their informed consent.
They then went through a role-play situation using a mixed-presence collaboration scenario where they were divided into two groups of 3. Both groups had to collaborate from two separate (and distant) WSDs, with synchronized views and connected with a standard audio-video link. The same participants then took part in a brainstorming session in order to identify and discuss problems related to awareness as well as potential solutions. 


\subsubsection{Participants}
In Q1 
2023, seven participants (5 females and 2 males) took part to this focus group. All of them were researchers in computer science or collaboration, and are used to working together in collaborative sessions, both in collocated and in mixed-presence settings, using conventional videoconferencing and collaborative document editing tools. One participant took part online as an observer, i.e. was connected through a videoconferencing tool on a PC.

\subsubsection{Role-play.} \label{sec:roleplay}

The mixed-presence role-play activity participants went through was derived from a collocated collaborative decision-making scenario adapted to WSDs~\cite{maquil2023establishing}. 
The scenario consists of decision-making activities to manage the medical supply chain in a hospital during a pandemic. Participants must first decide which personal equipment (gloves, masks, etc.) to buy for the medical staff, and then choose a supplier and delivery route.
Even though the scenario is based on a real use-case and was developed with medical staff and supply chain experts, it was adapted to be understandable and usable by non-experts.
The scenario incrementally presents different types of data to be analysed, in a predefined order that unfolds based on validation steps. Available actions include modifying values, panning, and zooming on a map, as well as selecting data and items. To induce collaboration, each participant plays the role of a key responsible person in this kind of decision-making (either the CEO, the head of ICU, the head of finance or the head of logistics). Each role receives additional information and constraints to help make the right decision.

%

The scenario was instantiated on two physically-distributed and touch-enabled WSDs, connected through an audio-video link that we consider as a baseline standard setup.
The first WSD (shown on the left of Figure~\ref{fig:teaser}) is the 360$^{\circ}$ Immersive Arena, a curved display reaching 2m in height that is composed of 12 screens (in portrait mode, 4K resolution each), with a diameter of $3.64$m. Eight of the screens 
were used as surfaces for the study. 
The second screen setup is a flat and vertical WSD, composed of two 4K touchscreens (in landscape mode, 65 inches each) with a total resolution of $7680\times2160$ pixels for a surface of $3$m$\times0.9$m (see Figure~\ref{fig:teaser} on the right). 
While the screen configurations of our WSDs are dissimilar, we argue that remote collaboration systems in practice are often also heterogeneous.
Both systems were placed in separate rooms. 
The video link was implemented with a single camera positioned in the middle of each WSD, filming the participants from the front. The raw video feed is shown to the remote users through a window placed in the middle of their WSD (yellow dotted rectangles in Figure \ref{fig:teaser}). 
A microphone was also placed in the middle of each WSD and is used to send the recorded audio to a speaker on the remote side. 
The groups successfully completed the scenario tasks in 20 minutes.


\subsubsection{Brainstorming.}
\label{sec:brainstorming}
To animate the brainstorming, we used the Lotus blossom technique~\cite{voehl2016lotus}. The aim of this creative-innovative-thinking technique is to foster the generation of a large number of ideas and solutions by decomposing the initial problem into sub-parts, organised into a matrix. 
The participants were brought together and reorganised into three groups of two (balanced such that each group was composed of one participant from each WSD). The online participant was working on their own. First, each group filled the Lotus blossom matrix. They then presented and explained the problems and solutions that they found, so that other groups could give their opinion and discuss these elements. 

\subsubsection{Data collection and analysis}
The role-play situation and the brainstorming were audio recorded.  
We aggregated the reported problems, needs, and potential solutions into a list, through analysis of the recorded audio and the lotus blossom matrices. We then categorised each item according to the element(s) of WA they mainly address. The results, including 
descriptions and illustrative diagrams, 
are available online\footnote{
Temporarily added as a PDF in Appendix A during the review process.
}.

\section{Results} \label{sec:results}

\subsection{Focus group results}
During the brainstorming, participants shared in total 28 ideas. We discarded 8 of the ideas (red rows in the spreadsheet) as they were not related to WA but rather suggested new features for the application (e.g. annotations, subtitles and automatic translations) or were related to technical issues (e.g. audio quality and volume). 
The remaining 20 ideas are grouped according to the four topics, as defined by participants in their lotus blossom matrices: \textit{Audio}, \textit{Video}, \textit{Pointing}, and \textit{Attention}. 
We do not count the number of times an idea was enunciated as ideas raised by one group were generally not repeated by subsequent groups.

\subsubsection{Audio}
To notify the remote side that someone is willing to speak, participants suggested a \textit{hand raised} feature similar to those from existing videoconferencing systems, possibly augmented with notification sounds, e.g. \textit{``a sort of beep''} (P1).

Local participants listening to remote ones indicated they sometimes had issues identifying what the remote side was referring to. They therefore suggested using spatialised audio, i.e. emitting audio from the speaker that is closest to the area of interest.
Another option that was envisioned to tackle the problem was the automatic highlighting of parts of the displayed content based on items that would be mentioned by the remote speaker (e.g. through keyword detection).

The participants knew each other's voice and therefore had no trouble identifying who was the remote speaker. However, they still reflected on issues they sometimes encounter in remote meetings involving less familiar participants, where it can be difficult to know who is talking. So they suggested that the existing video feed showing remote participants should zoom in on the speaker.

\subsubsection{Video}
More specifically on the topic of the video feed, participants pointed out that remote collaborators would sometimes get out of view and therefore suggested a broader field of view for the corresponding camera, or delimiting a zone on the floor within which users should stay to remain visible to the remote collaborators. 
Furthermore, participants said the video feed sometimes felt too small as they could not always see their collaborators well.  
In terms of placement, participants indicated that the default middle position was sometimes out of sight when they were focusing on other content. They would like this window to remain visible no matter which part of the screen they are looking at. They therefore suggested making it possible to move the video stream manually, or even \textit{``automatically move to the screen which is currently discussed''} (P2), as otherwise users \textit{``would not [manually] do it, probably''} (P3). The participants also mentioned that it was \textit{``good to have the video feedback close to the camera to have some sort of eye contact''} (P4). 

Participants also requested the ability \textit{``to see what the others were seeing''} (P4) on the video transmission, as they wanted to know whether they were within the camera's field of view, so that \textit{``each group sees both locations''} (P5). They actually had access to the recordings of the local camera but that local feed was likely too small for them to notice (see green rectangle on Figure \ref{fig:teaser}).

Finally, some participants mentioned the prospect of having a camera filming users from the back or the top, that could possibly zoom on arms or even hands to \textit{``to see what [remote collaborators] were doing''} (P6) and help with deictic references, in order to gather information such as \textit{``which screen [remote collaborators] are pointing [to]''} (P7).

\subsubsection{Pointing}
Participants pointed out that they could not see the hands and interactions of remote collaborators, so providing visual indicators for remote pointing occurrences was necessary. They suggested identifying cursors, e.g. through specific colours or shapes depending on the person, or role in the scenario, it originates from. 
Participants also reflected on other ways to identify cursors: with pictures or avatars of the participants or their roles, or simply by accompanying the cursors with labels containing the corresponding participants' initials.

In the discussion, participants anticipated the potential presence of too many cursors that would be distracting to the task at hand. They suggested that local cursors should be constantly but discreetly displayed to provide feedback on \textit{``whether [the system] is tracking you correctly''} (P4), and that remote cursors should be more salient but only appear when activated by remote participants (e.g. by speech detection, gesture recognition). 

The users would also like to distinguish between pointing towards screens, users, or possibly other physical items. 
Another pointing-related cue that was deemed useful is the need for post-interaction feedback indicating who did what.

\subsubsection{Attention}
Participants suggested to monitor gaze to know whether remote collaborators are looking at a screen or discussing privately, but not too precisely as this might feel intrusive and possibly be inaccurate anyway. 
Attention tracking was pointed out as a way to identify attention-pointing mismatches, i.e. \textit{``when you are looking at the screen and the other group will point on a different screen''} (P3). This also applies to attention-action mismatches, when users miss the action (e.g. changing a value) performed by a remote collaborator.

\subsection{Analysis from the perspective of WA}
To better understand how the proposed ideas relate to WA, we categorised the 20 retained ideas according to elements of WA, as proposed by
Gutwin et al.~\cite{gutwin2002descriptive}.


As part of that process, we found that three of the categories were unclear in the context of WSDs and adjusted the questions accordingly. More specifically, 
the categories \textit{View}, and \textit{Reach} could either define what users can view and reach from the current position in the room, or what they are able to view and reach in their site. Since in WSD environments, users can easily move around to view and reach screen content, we opted for the second variation and therefore defined \textit{View} and \textit{Reach} as what is available to be viewed and manipulated from a site. 
Regarding \textit{Location}, we found that the \textit{where} could refer to the actual position of a user in the workspace, or to a part of the WSD that a user is working on. Considering that \textit{Artefact} already covers WA information about the part of the WSD that a user is working on, we decided to define \textit{Location} as the physical position of the users in the environment. Artefact can then refer to a site in general (remote or local) or to a more precise position in relation to the screens. 

The categorisation of the comments in regards of the updated definitions shows that the vast majority of comments are related to \textit{Actions} (17/20), suggesting the importance of this type of WA information in a WSD setting (see Table~\ref{tab:framework_v2}). We can note that the actions that users mention are of different nature and can be very short: they cover manipulations of items on the screen, pointing gestures during discussions, and more subtle moves indicating to whom speech is addressed or whether there is an agreement. Furthermore, many of the mentioned actions are inherently bound with a location (e.g. pointing towards an item on the screen), so there is a high number of comments (8/20) on \textit{Artefact} as well. We also note that some of the actions are bound to a collaborator (e.g. agreeing), which explains the number of comments related to \textit{Authorship} (5/20).  

We can further see that \textit{Gaze} and \textit{View} were each only mentioned once (1/20). In contrast to \textit{Authorship} or \textit{Artefacts}, these are not directly linked to an action, but are required more generally, independently of what action is being done. This may explain why participants rarely mention them. We also note that participants could see both WSDs during the introduction to the session and were told that the data being displayed on both sides would be synchronised. This mitigated the need for \textit{View} awareness cues specifically.
On the topic of \textit{Presence}, the comments (5/20) are mainly dealing with the fact that collaborators could not always be seen on the video stream. 

There were no comments related to \textit{Identity}, \textit{Intention}, \textit{Location} and \textit{Reach}. The absence of comments in these categories can be explained by the context of the scenario. Since the role-play was part of a face-to-face meeting and everyone knew each other, there were no problems related to knowing who was in the workspace, their identity, and from where they were working. Furthermore, there were no questions related to \textit{Reach}, as both sites knew they had a similar setup with the same manipulation possibilities. Finally, we assume that \textit{Intention} was not a problem because the scenario was driven by tasks which indicate the type of actions that are to be expected at each stage. The audio link between both sites also helped clarify intentions through verbal communication when needed.

\section{Discussion}
The categorisation has shown that one WA cue idea frequently addresses several WA elements simultaneously. For example, cursors showing pointing gestures do not only provide information on \textit{Action}, but also on \textit{Artefact} and ideally \textit{Authorship}. 
We noticed that, depending on the type of WA information, the same combinations of WA elements were addressed. Therefore, we adjusted the framework to introduce new categories that each are related to a different WA component and include a distinct set of WA elements, as seen in Table \ref{tab:framework_v2}.  

WA information related to the \textit{Environment} informs on the WSD setup, i.e. who is participating to the session, what is displayed on the screens, what can be modified. Such information would normally remain identical throughout a session, but might also change sporadically. Therefore, \textit{Environment} awareness cues (e.g. a list of participants) could be shown on a fixed location where they remain visible throughout the entire session, so that people can refer to them as needed.

WA information related to \textit{Actions} is highly spatial, dynamic, and rich. Depending on the scenario, several \textit{Action} cues may be needed, and their design should be adapted to the action itself (e.g. pointing, explaining, agreeing, manipulating the interface, etc.). They also often require \textit{Authorship} and \textit{Artefact} information alongside them to be correctly understood. The particular challenge about the design of \textit{Action} awareness cues is to have them noticeable, understandable and clear, but not disturbing, distracting or overwhelming.

Finally, WA information related to \textit{Attention} is needed throughout the whole session, independently from the type of action. Users would like to see who is paying attention and whether remote participants are currently talking to each other or following the \textit{Actions} of the local team. Since they are naturally based on gaze, \textit{Attention} awareness cues might feel intrusive to the user. Special caution is therefore needed to provide the required information while respecting the privacy of users. 

\begin{table}
  \caption{Components of awareness information, including the related WA elements and refined questions. The * marking denotes differences with the original table.}
  \label{tab:framework_v2}
  \begin{tabular}{p{0.39\linewidth} | p{0.13\linewidth} | p{0.47\linewidth}} 
    \hline
    \textbf{WA Component}&\textbf{WA \mbox{Element}}&\textbf{Question}\\
    \hline
    \textbf{Environment}  
            &Presence   &Is anyone in the workspace? (5)\\
    \multirow{2}{\linewidth}{Configuration of the WSD environment(s).}
            &Identity   &Who is participating? Who is that? (0)\\
            &View*      &What is available to be viewed? (1)\\
            &Reach*     &What is available to be manipulated? (0)\\
    \hline
    \textbf{Action}  
            &Action       &What are they doing? (17)\\
    \multirow{3}{\linewidth}{Actions done as part of the collaborative  work. Here: pointing, speaking, manipulating, asking to speak, (dis)agreeing.}
            &Authorship   &Who is doing that? (5)\\
   
            &Artefact     &What object are they working on? (8)\\
     
            &Location*    &From which location are they working? (0)\\
    \hline
    \textbf{Attention}  
            &Gaze       &Where are they looking? (1)\\
    State of others’ attention. & & \\
    \hline
\end{tabular}
\end{table}

\section{Conclusion}

In this paper, we conducted a focus group consisting of a role-play of a mixed-presence decision-making scenario and a brainstorming on the issues participants met. We analysed the gathered comments using 
Gutwin et al.'s workspace awareness framework and proposed an adjusted version of that framework, suitable for WSDs. The main contributions of the presented research are:
\begin{itemize}
    \item We found that support for WA on \textit{Action}, \textit{Artefact}, \textit{Authorship} and \textit{Presence} is most crucially required, and appropriate WA cues therefore need to be created for collaborative mixed-presence sessions on WSDs.
    \item We have shown that the WA information needed for this particular context can be structured in three different categories and are related to (i) the Environment, (ii) Actions and (iii) Attention.
    \item We argue that the upmost challenge in the design of WA cues is the balance between a noticeable design, while avoiding occlusion and disturbance, and preserving an acceptable level of privacy. 
\end{itemize}

A limitation of the present results is that they are tied to the nature of the task we chose. Indeed, results might have been different if the goals had required more manipulations, or if participants 
had been observed in situ, while collaborating as part of their day-to-day job activities.
In addition, the fact that participants knew each other well may have had an effect on their perception of problems during the scenario and possibly explains that only few comments on \textit{Authorship} were raised, and none regarding \textit{Identity} .
Their previous experience with WSDs and mixed-presence meetings could also have played a role.

In future work, we will use the proposed framework to design, implement and evaluate an initial set of cues. Furthermore, we will explore variations of designs for individual WA cues types and test them to investigate how they improve the WA of mixed-presence collaborators and how they impact their work. 
This would help continue the validation of the proposed adjusted framework and support the understanding of WA in mixed-presence collaboration settings with WSDs, so that WA cues may be properly designed in that context.

\begin{credits}
\subsubsection{\ackname} 
Authors would like to thank all participants to the focus group.

\subsubsection{\discintname}
\end{credits}


\bibliographystyle{splncs04}

\clearpage
\thispagestyle{empty}
\section*{Appendix A: brainstorming results}
\label{apx:results}
\includegraphics[trim=2.3cm 0 0 0,scale=0.75,page=1]{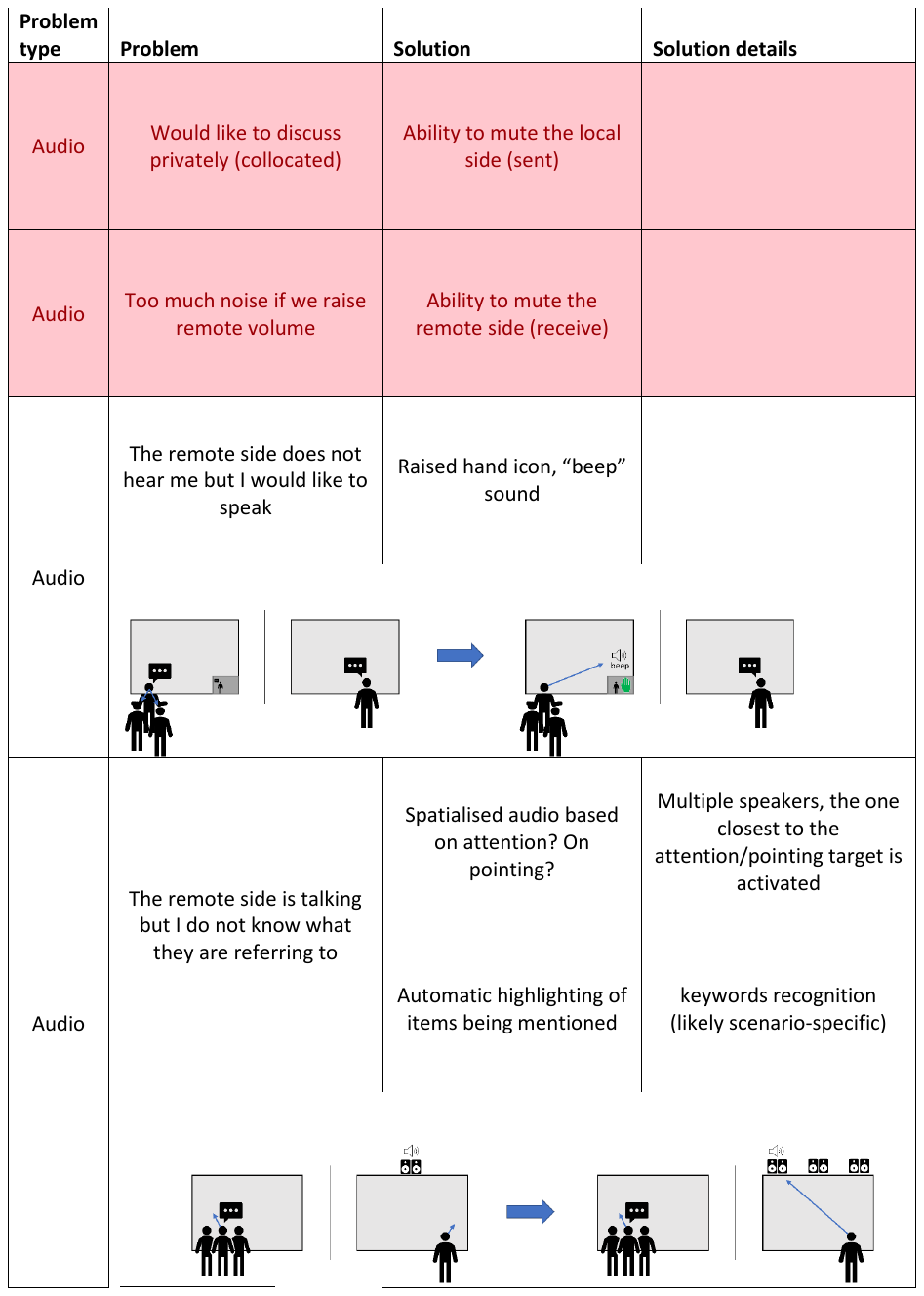}
\newpage
\includegraphics[trim=3cm 0 0 0,scale=0.75,page=2]{cdve-tmp-apx.pdf}
\newpage
\includegraphics[trim=3cm 0 0 0,scale=0.75,page=3]{cdve-tmp-apx.pdf}
\newpage
\includegraphics[trim=3cm 0 0 0,scale=0.75,page=4]{cdve-tmp-apx.pdf}
\newpage
\includegraphics[trim=3cm 0 0 0,scale=0.75,page=5]{cdve-tmp-apx.pdf}
\newpage
\includegraphics[trim=3cm 0 0 0,scale=0.75,page=6]{cdve-tmp-apx.pdf}

\end{document}